\documentclass[aps,prd,reprint,showpacs,superscriptaddress,twocolumn]{revtex4-1}
\usepackage{amsfonts}
\usepackage{graphicx}
\usepackage{amsmath}
\usepackage{amssymb}
\usepackage[breaklinks=true,dvips]{hyperref}
\usepackage{breakurl}
\usepackage[utf8x]{inputenc}
\usepackage{mathrsfs}
\usepackage{color}

\hypersetup{
  colorlinks=true,linkcolor=blue,citecolor=blue,
  filecolor=blue,urlcolor=blue,breaklinks=true
}
\begin{document}

\title{Singularity transition in a Yukawa-Reissner-Nordstr\"om spacetime}
\author{Thiago Prud\^encio}
\affiliation{Universidade Federal do Maranh\~ao - UFMA, Curso Interdisciplinar em Ci\^encia e Tecnologia - CCCT, 
Campus Bacanga, 65080-805, Sao Lu\'is-MA, Brazil.}


\begin{abstract}
A screened charge by a Yukawa potential implies that the corresponding electric field is assimptotically zero far away from a small radius 
scaling. Considering a screened charge, we have a modified metric of type 
Yukawa-Reissner-Nordstr\"om spacetime, that assimptotically decays from a Reissner-Nordstr\"om spacetime 
to a Schwarzschild spacetime. This implies that the singularity is screened by distance reducing from two singularities to one single singularity 
for a suitable distance. We consider the tetrad fields associated to the Yukawa-Reissner-Nordstr\"om spacetime and show that a linear motion in the 
proper time lead to suitable estimatives to the Yukawa screening parameter associated to a Schwarzschild radius.
\end{abstract}

\maketitle

\begin{widetext}

\section{Introduction}

In supersymmetric theories the mass of a given state is bounded below the values of some charges, in a Schwarzschild ($M\geq 0$), 
in a Reissner-Nordstr\"om
($M\geq |Q|$) corresponding to absence of naked singularities. Extreme dilaton black holes, with electric and magnetic charges, 
satisfy $M \geq \frac{1}{\sqrt{2}}(|Q| + |P|)$, corresponding to a supersymmetry bound of a $N=4,d=4$ supergravity, the Bekenstein-Hawking entropy 
is given by $S=2\pi |PQ|$. In its extremal limit, the thermal description breaks down and consequently there is no Hawking radiation,the black hole cannot continue to evatorate by 
emitting (uncharged) elementary particles, what would violate the supersymmetric positivity bound  \cite{kallosh1}. This is an important for the theories of last stages 
of black hole evaporation, in a Schwarzschild black hole, this occurs when the mass of the black hole achieves the Planck mass
\begin{eqnarray}
M_{p}=\sqrt{\hbar c/G}.
\end{eqnarray}
In the case of a Reissner-Nordstr\"om black hole, the evaporation stops when the Planck mass achieves the absolute value of its charge $M_{p}=|Q|$, for a
large charge, the last stages of evaporation occur for a mass much greater than Planck mass. In this case, however, the black hole can discharge by the 
creation of pairs of charged elementary particles \cite{wilczek}. In order to study only the quantum gravity effects, separated from pair productions, it is possible 
to consider the electric and magnetic fields not produced by charged particles, but originating from the singularity or from the infinity. In this case, by embedding 
the original bosonic theory in a supersymmetric theory, the extremal Reissner-Nordstr\"om black hole can be embedded in $N=2$ supergravity, vanishing 
any higher order quantum corrections if the theory is free from anomalies. Since a $N=2$ supergravity has a one-loop anomaly, a supersymmetric embedding 
of charged dilaton black holes is achieved in $d=4$, $N=4$ supergravity, where the anomalies can be cancelled. 

Extremal Bogomol'nyi-Prasad-Sommerfield (BPS) black holes in 
$\mathcal{N}=8$ supergravity have been associated to 
massive representations of a $\mathcal{N}=8$ supersymmetry algebra, with three categories according to the black hole background according to 
preservation of $1/2$, 
$1/4$ or $1/8$ from the original supersymmetry state \cite{ferrara,ferrara2}. An important point in the study of extremal black holes 
is the classification of BPS states preserving 
the supersymmetry states and its parallel to the groups and orbits of timelike, lightlike and spacelike vectors in Minkowski space \cite{ferrara}. 
The string-theoretic interpretation of black holes given in terms of $Dp$-branes wrapping around 
compactified dimensions associated to qubits from quantum information (QI) lead to the so-called black hole qubit correspondence (BHQC) 
\cite{borsten1,borsten11,borsten12,levay4,borsten4,levay3,levay5,levay6,borsten5,borsten2,levay7}. Important achievements have 
been realized in this context, associated with the black hole entropy emerging from the solution of $N = 2$ supergravity 
STU model of string compactification and tripartite entanglement measurement 
\cite{kallosh,duff}, black hole configurations in STU supergravity \cite{marrani1}, classifications of entanglement state \cite{borsten13}, identification of the Hilbert space for
qubits associated to the wrapped branes inside 
the cohomology of the extra dimensions \cite{levay,borsten12} and association to quantum circuits \cite{prudencio1}.

A Reissner-N\"ordstrom spacetime is an important generalization of a Schwarzschild metric in the presence of charge and absence of spin. This 
configuration is a solution of Einstein-Maxwell field equations, corresponding to the gravitational field caused by 
a charged non-rotating mass \cite{wilczek2}. 
An important aspect 
of this gravitational scenario is the absence of a single horizon by the presence of a double singularity \cite{naked,naked1,naked2,naked2}. In the 
tereparallel formalism \cite{vanessa1,vanessa2,vanessa3} the 
description of such a spacetime is appropriate the use of tetrad fields. When a screened charge is introduced via a 
Yukawa type singularity, the electric field is reduced to zero for large distances.

Here we consider a $(3+1)$-dimensional Yukawa-Reissner-Nordstr\"om spacetime described by the corresponding Reissner-Nordstr\"om metric with screened 
charge
\begin{eqnarray}
ds^{2}=\left(1-\frac{R_{S}}{r} + \frac{R_{0}^{2}}{r^{2}}e^{-\mu r} \right)c^{2}dt^{2} 
-\frac{1}{\left(1-\frac{R_{S}}{r} + \frac{R_{0}^{2}}{r^{2}}e^{-\mu r}\right)}dr^{2} -r^{2}d\theta^{2}-r^{2}\sin^{2}\theta d\varphi^{2},
\end{eqnarray} 
where the Schwarzschild radius in Schwarzschild metric, $R_{S}=2GM/c^{2}$ is complemented by a contribution of 
charge parameter $Q(r)=Q_{0}e^{-\mu r}$, with charge radius 
\begin{eqnarray}
R_{Q}(r)=R_{0}e^{-\mu r} &=& \sqrt{\frac{Q_{0}^{2}G}{4\pi\varepsilon_{0}c^{4}}}e^{-\mu r}.
\end{eqnarray}
and associated screened electric field is given by
\begin{eqnarray}
{\bf E}(r)&=& \frac{-Q_{0}e^{-\mu r}}{4\pi \varepsilon_{0}r^{2}}\hat{r}
\end{eqnarray}
The Reissner-Nordstr\"om classification of sigularities comes from roots of the equation 
\begin{eqnarray}
R^{2}_{\pm}(r)-R_{S}R_{\pm}(r) + R_{Q}^{2} = 0,
\end{eqnarray}
where the sigularities are given by
\begin{eqnarray}
R_{\pm}(r)= \frac{R_{S}}{2} \pm \sqrt{\left(\frac{R_{S}}{2}\right)^{2}-R_{0}^{2}e^{-2\mu r}}.
\end{eqnarray}
A consequence is that there is more than one singularity, whose origin is the presence of charge. The Schwarzschild radius can be 
rewritten in terms of 
\begin{eqnarray}
  R_{+}(r) + R_{-}(r) &=& R_{S}, 
\end{eqnarray}
In terms of these singularities the radius can be written 
\begin{eqnarray}
R_{Q}(r)&=&\sqrt{R_{+}(r)R_{-}(r)}.
\end{eqnarray}
Consequently
\begin{eqnarray}
R_{+}(r)R_{-}(r)&=& R_{0}^{2}e^{-2\mu r}.
\end{eqnarray}

The metric can be rewritten as
\begin{eqnarray}
ds^{2}=\frac{(r-R_{+}(r))(r-R_{-}(r))}{r^{2}}c^{2}dt^{2} 
-\frac{r^{2}}{(r-R_{+}(r))(r-R_{-}(r))}dr^{2} -r^{2}d\theta^{2}-r^{2}\sin^{2}\theta d\varphi^{2}.
\end{eqnarray}  
or in a explicit way, the Yukawa-Reissner-Nordstr\"om spacetime is given by
\begin{eqnarray}
ds^{2}&=& \frac{\left(r-\left[\frac{R_{S}}{2} + \sqrt{\left(\frac{R_{S}}{2}\right)^{2}-R_{0}^{2}e^{-2\mu r}}\right]\right)\left(
r-\left[\frac{R_{S}}{2} - \sqrt{\left(\frac{R_{S}}{2}\right)^{2}-R_{0}^{2}e^{-2\mu r}}\right]\right)}{r^{2}}c^{2}dt^{2} \nonumber \\
&-&\frac{r^{2}}{\left(r-\left[\frac{R_{S}}{2} + \sqrt{\left(\frac{R_{S}}{2}\right)^{2}-R_{0}^{2}e^{-2\mu r}}\right]\right)\left(r
-\left[\frac{R_{S}}{2} - \sqrt{\left(\frac{R_{S}}{2}\right)^{2}-R_{0}^{2}e^{-2\mu r}}\right]\right)}dr^{2} \nonumber \\
&-& r^{2}d\theta^{2}-r^{2}\sin^{2}\theta d\varphi^{2},
\end{eqnarray}  
Notice that as $r\rightarrow \infty$, the spacetime is driven assimptotically to a Schwarzschild spacetime.

\begin{figure}[h]
\centering
\includegraphics[scale=0.35]{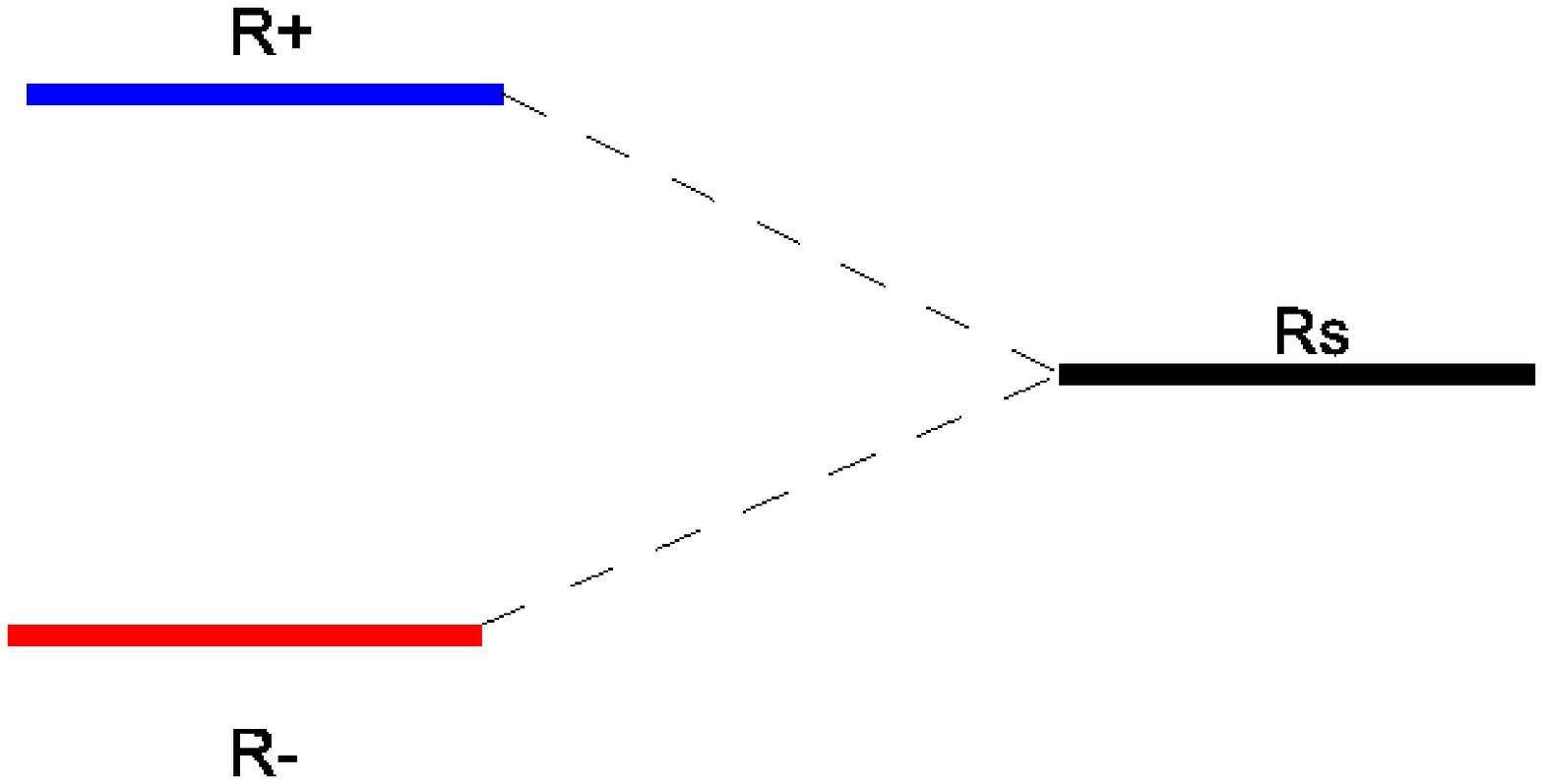}
\includegraphics[scale=0.35]{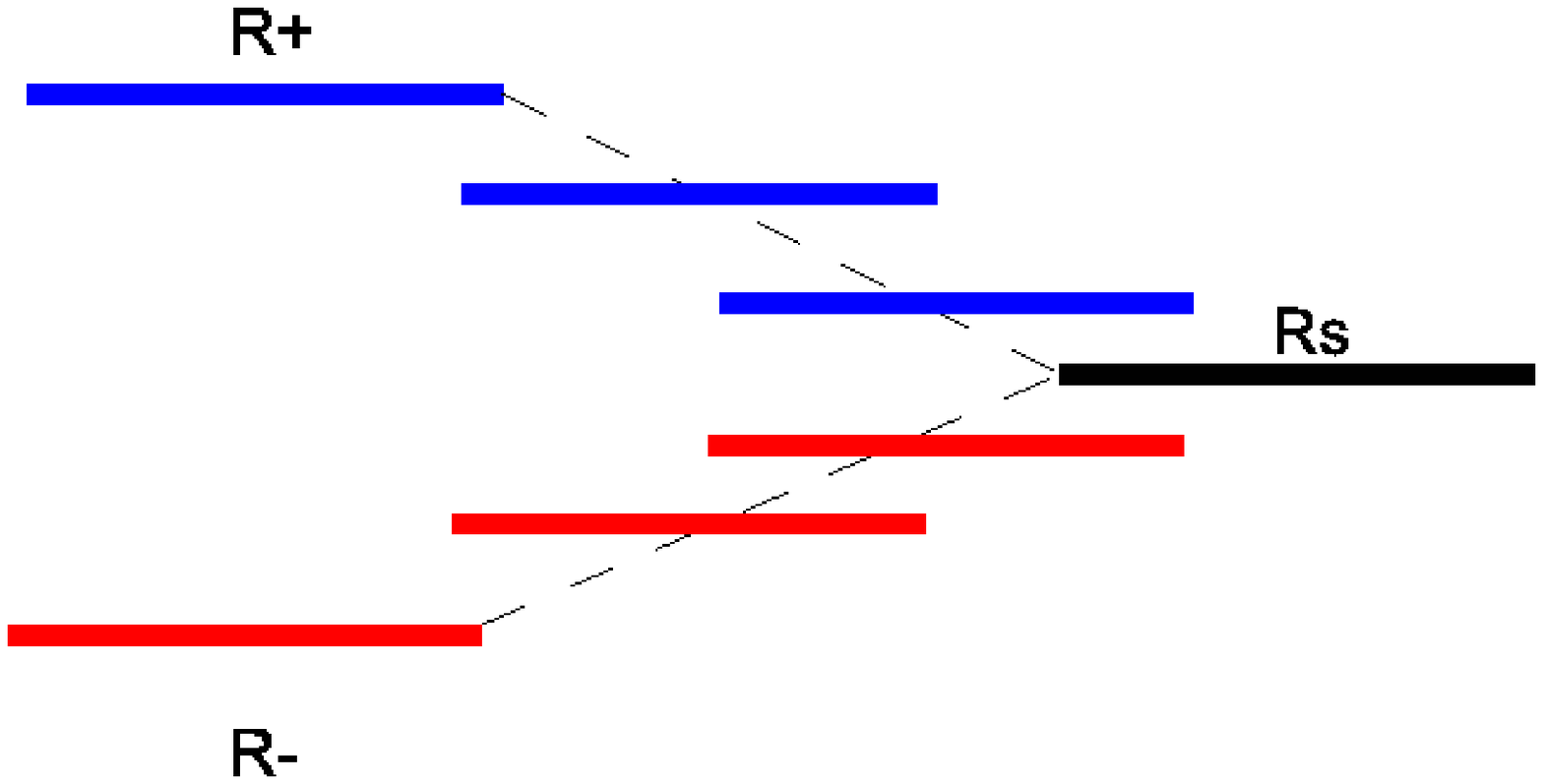}
\caption{(Color online) Description of the double singularity and the transition to a single singularity due to Yukawa screening.}
\label{b1}
\end{figure}

\section{Yukawa-Reissner-Nordstr\"om tetrad fields}

We can write the metric in terms of tetrads fields $e^{a}_{\mu}(x)$, relating the curved spacetime metric to the flat matric by means 
of the relation $g_{\mu\nu}=e^{a}_{\mu}(x)e^{b}_{\nu}(x)\eta_{ab}$, where $(\eta_{ab})=(\eta^{ab})=diag(+---)$, $\eta_{ab}\eta^{ac}=\delta^{c}_{b}$ and $e^{a}_{\mu}(x)$ satisfy
$e_{a}^{\mu}(x)e^{b}_{\mu}(x)=\eta_{ab}$, $e^{a}_{\mu}(x)= g^{\mu\nu}e^{a}_{\mu}(x)$, 
$e^{\mu}_{a}(x)=g_{\mu\nu}e^{\nu}_{a}(x)$.
We metric can then be rewritten in the following manner
\begin{eqnarray}
ds^{2}
= e^{a}_{\mu}(x)e^{b}_{\nu}(x)\eta_{ab}dx^{\mu}dx^{\nu}
\end{eqnarray}
In this form, we can write the metric in terms of a non-coordinate basis $\hat{\Theta}^{a}=e^{a}_{\mu}(x)dx^{\mu}$,
as follows $ds^{2}= \eta_{ab}e^{a}_{\mu}(x)dx^{\mu}e^{b}_{\nu}(x)dx^{\nu}=\eta_{ab}\hat{\Theta}^{a}\hat{\Theta}^{b}$. We now have
\begin{eqnarray}
ds^{2}= \hat{\Theta}^{0}\hat{\Theta}^{0} - \hat{\Theta}^{1}\hat{\Theta}^{1} - \hat{\Theta}^{2}\hat{\Theta}^{2} - \hat{\Theta}^{3}\hat{\Theta}^{3}
\end{eqnarray}
The correspondence with the differential terms in the metric will lead to
\begin{eqnarray}
\hat{\Theta}^{0}(r)\hat{\Theta}^{0}(r)&=& \frac{\left(r-\left[\frac{R_{S}}{2} + \sqrt{\left(\frac{R_{S}}{2}\right)^{2}-R_{0}^{2}e^{-2\mu r}}\right]\right)\left(
r-\left[\frac{R_{S}}{2} - \sqrt{\left(\frac{R_{S}}{2}\right)^{2}-R_{0}^{2}e^{-2\mu r}}\right]\right)}{r^{2}}c^{2}dt^{2},\\
\hat{\Theta}^{1}(r)\hat{\Theta}^{1}(r)&=& \frac{r^{2}}{\left(r-\left[\frac{R_{S}}{2} + \sqrt{\left(\frac{R_{S}}{2}\right)^{2}-R_{0}^{2}e^{-2\mu r}}\right]\right)\left(
r-\left[\frac{R_{S}}{2} - \sqrt{\left(\frac{R_{S}}{2}\right)^{2}-R_{0}^{2}e^{-2\mu r}}\right]\right)}dr^{2} \\
\hat{\Theta}^{2}(r)\hat{\Theta}^{2}(r)&=& r^{2}d\theta^{2}\\
\hat{\Theta}^{3}(r,\theta)\hat{\Theta}^{3}(r,\theta)&=& r^{2}\sin^{2}\theta d\varphi^{2}
\end{eqnarray}
Consequently, the non-vanishing contributions are given by
\begin{eqnarray}
\hat{\Theta}^{0}(r)&=& 
\frac{\sqrt{\left(r-\left[\frac{R_{S}}{2} + \sqrt{\left(\frac{R_{S}}{2}\right)^{2}-R_{0}^{2}e^{-2\mu r}}\right]\right)\left(
r-\left[\frac{R_{S}}{2} - \sqrt{\left(\frac{R_{S}}{2}\right)^{2}-R_{0}^{2}e^{-2\mu r}}\right]\right)}}{r}cdt, \label{lkoi9871}\\
\hat{\Theta}^{1}(r)&=& 
\frac{r}{\sqrt{\left(r-\left[\frac{R_{S}}{2} + \sqrt{\left(\frac{R_{S}}{2}\right)^{2}-R_{0}^{2}e^{-2\mu r}}\right]\right)\left(
r-\left[\frac{R_{S}}{2} - \sqrt{\left(\frac{R_{S}}{2}\right)^{2}-R_{0}^{2}e^{-2\mu r}}\right]\right)}}dr \\
\hat{\Theta}^{2}(r)&=& 
rd\theta \\
\hat{\Theta}^{3}(r,\theta)&=& 
r\sin\theta d\varphi. \label{lkoi98712}
\end{eqnarray}
The associated non-zero tetrad fields and their derivatives are given by 
\begin{eqnarray}
e^{0}_{0}(r)&=& \frac{\sqrt{\left(r-\left[\frac{R_{S}}{2} + \sqrt{\left(\frac{R_{S}}{2}\right)^{2}-R_{0}^{2}e^{-2\mu r}}\right]\right)\left(
r-\left[\frac{R_{S}}{2} - \sqrt{\left(\frac{R_{S}}{2}\right)^{2}-R_{0}^{2}e^{-2\mu r}}\right]\right)}}{r}, 
\\
e^{1}_{1}(r)&=& \frac{r}{\sqrt{\left(r-\left[\frac{R_{S}}{2} + \sqrt{\left(\frac{R_{S}}{2}\right)^{2}-R_{0}^{2}e^{-2\mu r}}\right]\right)\left(
r-\left[\frac{R_{S}}{2} - \sqrt{\left(\frac{R_{S}}{2}\right)^{2}-R_{0}^{2}e^{-2\mu r}}\right]\right)}}, 
\\
e^{2}_{2}(r)&=& r, 
\\
e^{3}_{3}(r,\theta)&=& r\sin\theta,
\end{eqnarray}
\begin{eqnarray}
\frac{de^{0}_{0}(r)}{dr}
&=& -\frac{\sqrt{\left(r-\left[\frac{R_{S}}{2} + \sqrt{\left(\frac{R_{S}}{2}\right)^{2}-R_{0}^{2}e^{-2\mu r}}\right]\right)\left(
r-\left[\frac{R_{S}}{2} - \sqrt{\left(\frac{R_{S}}{2}\right)^{2}-R_{0}^{2}e^{-2\mu r}}\right]\right)}}{r^{2}} \nonumber \\
&+& \frac{\left(1-\frac{\mu R_{0}^{2}e^{-2\mu r}}{\sqrt{\left(\frac{R_{S}}{2}\right)^{2}-R_{0}^{2}e^{-2\mu r}}}\right)\left(
r-\left[\frac{R_{S}}{2} - \sqrt{\left(\frac{R_{S}}{2}\right)^{2}-R_{0}^{2}e^{-2\mu r}}\right]\right)}{2r\sqrt{\left(r-\left[\frac{R_{S}}{2} + \sqrt{\left(\frac{R_{S}}{2}\right)^{2}-R_{0}^{2}e^{-2\mu r}}\right]\right)\left(
r-\left[\frac{R_{S}}{2} - \sqrt{\left(\frac{R_{S}}{2}\right)^{2}-R_{0}^{2}e^{-2\mu r}}\right]\right)}} \nonumber \\
&+& \frac{\left(r-\left[\frac{R_{S}}{2} + \sqrt{\left(\frac{R_{S}}{2}\right)^{2}-R_{0}^{2}e^{-2\mu r}}\right]\right)\left(
1 + \frac{\mu R_{0}^{2}e^{-2\mu r}}{\sqrt{\left(\frac{R_{S}}{2}\right)^{2}-R_{0}^{2}e^{-2\mu r}}}\right)}{2r\sqrt{\left(r-\left[\frac{R_{S}}{2} 
+ \sqrt{\left(\frac{R_{S}}{2}\right)^{2}-R_{0}^{2}e^{-2\mu r}}\right]\right)\left(
r-\left[\frac{R_{S}}{2} - \sqrt{\left(\frac{R_{S}}{2}\right)^{2}-R_{0}^{2}e^{-2\mu r}}\right]\right)}}
 \nonumber \\
&&
\end{eqnarray}
\begin{eqnarray}
\frac{de^{1}_{1}(r)}{dr}
&=& \frac{1}{\sqrt{\left(r-\left[\frac{R_{S}}{2} 
+ \sqrt{\left(\frac{R_{S}}{2}\right)^{2}-R_{0}^{2}e^{-2\mu r}}\right]\right)\left(
r-\left[\frac{R_{S}}{2} - \sqrt{\left(\frac{R_{S}}{2}\right)^{2}-R_{0}^{2}e^{-2\mu r}}\right]\right)}}, \nonumber\\
&-& \frac{1}{2} \frac{r\left(1-\frac{\mu R_{0}^{2}e^{-2\mu r}}{\sqrt{\left(\frac{R_{S}}{2}\right)^{2}-R_{0}^{2}e^{-2\mu r}}}\right)\left(
r-\left[\frac{R_{S}}{2} - \sqrt{\left(\frac{R_{S}}{2}\right)^{2}-R_{0}^{2}e^{-2\mu r}}\right]\right)}{\left[\sqrt{\left(r-\left[\frac{R_{S}}{2} 
+ \sqrt{\left(\frac{R_{S}}{2}\right)^{2}-R_{0}^{2}e^{-2\mu r}}\right]\right)\left(
r-\left[\frac{R_{S}}{2} - \sqrt{\left(\frac{R_{S}}{2}\right)^{2}-R_{0}^{2}e^{-2\mu r}}\right]\right)} \right]^{3}}, \nonumber\\
&-& \frac{1}{2} \frac{r\left(r-\left[\frac{R_{S}}{2} + \sqrt{\left(\frac{R_{S}}{2}\right)^{2}-R_{0}^{2}e^{-2\mu r}}\right]\right)\left(
1 + \frac{\mu R_{0}^{2}e^{-2\mu r}}{\sqrt{\left(\frac{R_{S}}{2}\right)^{2}-R_{0}^{2}e^{-2\mu r}}}\right)}{\left[\sqrt{\left(r-\left[\frac{R_{S}}{2} 
+ \sqrt{\left(\frac{R_{S}}{2}\right)^{2}-R_{0}^{2}e^{-2\mu r}}\right]\right)\left(
r-\left[\frac{R_{S}}{2} - \sqrt{\left(\frac{R_{S}}{2}\right)^{2}-R_{0}^{2}e^{-2\mu r}}\right]\right)} \right]^{3}}, \nonumber\\
\end{eqnarray}
\begin{eqnarray}
\frac{de^{2}_{2}(r)}{dr}&=& 1,
\end{eqnarray}
\begin{eqnarray}
\frac{\partial e^{3}_{3}(r,\theta)}{\partial r}&=& \sin\theta,  
\end{eqnarray}
\begin{eqnarray}
\frac{\partial e^{3}_{3}(r,\theta)}{\partial \theta }&=& r\cos\theta.
\end{eqnarray}
Taking the exterior derivatives,
$d\hat{\Theta}^{a}= \partial_{\mu}{e}^{a}_{\nu}(x)dx^{\mu}\wedge dx^{\nu}$, explicitly, we have
\begin{eqnarray}
d\hat{\Theta}^{0} &=& -\frac{\sqrt{\left(r-\left[\frac{R_{S}}{2} + \sqrt{\left(\frac{R_{S}}{2}\right)^{2}-R_{0}^{2}e^{-2\mu r}}\right]\right)\left(
r-\left[\frac{R_{S}}{2} - \sqrt{\left(\frac{R_{S}}{2}\right)^{2}-R_{0}^{2}e^{-2\mu r}}\right]\right)}}{r^{2}}dr\wedge cdt  \nonumber \\
&+& \frac{\left(1-\frac{\mu R_{0}^{2}e^{-2\mu r}}{\sqrt{\left(\frac{R_{S}}{2}\right)^{2}-R_{0}^{2}e^{-2\mu r}}}\right)\left(
r-\left[\frac{R_{S}}{2} - \sqrt{\left(\frac{R_{S}}{2}\right)^{2}-R_{0}^{2}e^{-2\mu r}}\right]\right)}{2r\sqrt{\left(r-\left[\frac{R_{S}}{2} + \sqrt{\left(\frac{R_{S}}{2}\right)^{2}-R_{0}^{2}e^{-2\mu r}}\right]\right)\left(
r-\left[\frac{R_{S}}{2} - \sqrt{\left(\frac{R_{S}}{2}\right)^{2}-R_{0}^{2}e^{-2\mu r}}\right]\right)}}dr\wedge cdt  \nonumber \\
&+& \frac{\left(r-\left[\frac{R_{S}}{2} + \sqrt{\left(\frac{R_{S}}{2}\right)^{2}-R_{0}^{2}e^{-2\mu r}}\right]\right)\left(
1 + \frac{\mu R_{0}^{2}e^{-2\mu r}}{\sqrt{\left(\frac{R_{S}}{2}\right)^{2}-R_{0}^{2}e^{-2\mu r}}}\right)}{2r\sqrt{\left(r-\left[\frac{R_{S}}{2} 
+ \sqrt{\left(\frac{R_{S}}{2}\right)^{2}-R_{0}^{2}e^{-2\mu r}}\right]\right)\left(
r-\left[\frac{R_{S}}{2} - \sqrt{\left(\frac{R_{S}}{2}\right)^{2}-R_{0}^{2}e^{-2\mu r}}\right]\right)}}dr\wedge cdt 
 \nonumber \\
&&\\
d\hat{\Theta}^{1}&=& 0\\
d\hat{\Theta}^{2}
&=& dr\wedge d\theta \\
d\hat{\Theta}^{3}
&=& \sin\theta dr\wedge d\varphi + r\cos\theta d\theta\wedge d\varphi	\label{12kkj2}
\end{eqnarray}
As a consequence we can use the Maurer-Cartan structure equations in absence of torsion, $d\hat{\Theta}^{a} + \omega_{\mu b}^{a}(x)dx^{\mu}\wedge \hat{\Theta}^{b}= 0$ and 
write explicitly in this case
, 
\begin{eqnarray}
d\hat{\Theta}^{0} + \omega_{\mu 0}^{0}(x)dx^{\mu}\wedge \hat{\Theta}^{0} + \omega_{\mu 1}^{0}(x)dx^{\mu}\wedge \hat{\Theta}^{1} 
+ \omega_{\mu 2}^{0}(x)dx^{\mu}\wedge \hat{\Theta}^{2}+ \omega_{\mu 3}^{0}(x)dx^{\mu}\wedge \hat{\Theta}^{3}  &=& 0, \\
d\hat{\Theta}^{1} + \omega_{\mu 0}^{1}(x)dx^{\mu}\wedge \hat{\Theta}^{0} + \omega_{\mu 1}^{1}(x)dx^{\mu}\wedge \hat{\Theta}^{1} 
+ \omega_{\mu 2}^{1}(x)dx^{\mu}\wedge \hat{\Theta}^{2}+ \omega_{\mu 3}^{1}(x)dx^{\mu}\wedge \hat{\Theta}^{3}  &=& 0, \\
d\hat{\Theta}^{2} + \omega_{\mu 0}^{2}(x)dx^{\mu}\wedge \hat{\Theta}^{0} + \omega_{\mu 1}^{2}(x)dx^{\mu}\wedge \hat{\Theta}^{1} 
+ \omega_{\mu 2}^{2}(x)dx^{\mu}\wedge \hat{\Theta}^{2}+ \omega_{\mu 3}^{2}(x)dx^{\mu}\wedge \hat{\Theta}^{3}  &=& 0, \\
d\hat{\Theta}^{3} + \omega_{\mu 0}^{3}(x)dx^{\mu}\wedge \hat{\Theta}^{0} + \omega_{\mu 1}^{3}(x)dx^{\mu}\wedge \hat{\Theta}^{1} 
+ \omega_{\mu 2}^{3}(x)dx^{\mu}\wedge \hat{\Theta}^{2}+ \omega_{\mu 3}^{3}(x)dx^{\mu}\wedge \hat{\Theta}^{3}  &=& 0.
\end{eqnarray}

\section{Linear motion in a proper time}

Considering a motion is a proper time, we can write
\begin{eqnarray}
c^{2}=\left(\frac{ds}{d\tau}\right)^{2} &=& \frac{\left(r-\left[\frac{R_{S}}{2} + \sqrt{\left(\frac{R_{S}}{2}\right)^{2}-R_{0}^{2}e^{-2\mu r}}\right]\right)\left(
r-\left[\frac{R_{S}}{2} - \sqrt{\left(\frac{R_{S}}{2}\right)^{2}-R_{0}^{2}e^{-2\mu r}}\right]\right)}{r^{2}}c^{2}\left(\frac{dt}{d\tau}\right)^{2} \nonumber \\
&-&\frac{r^{2}}{\left(r-\left[\frac{R_{S}}{2} + \sqrt{\left(\frac{R_{S}}{2}\right)^{2}-R_{0}^{2}e^{-2\mu r}}\right]\right)\left(r
-\left[\frac{R_{S}}{2} - \sqrt{\left(\frac{R_{S}}{2}\right)^{2}-R_{0}^{2}e^{-2\mu r}}\right]\right)}\left(\frac{dr}{d\tau}\right)^{2} \nonumber \\
&-& r^{2}\left(\frac{d\theta}{d\tau}\right)^{2}-r^{2}\sin^{2}\theta \left(\frac{d\varphi}{d\tau}\right)^{2},  \label{89uop1}
\end{eqnarray}  
We can choose a linear motion where the angular velocities $\varphi$ and $\theta$ are constant
\begin{eqnarray}
\frac{d\varphi}{d\tau}&=&0, \\
\frac{d\theta}{d\tau}&=&0. 
\end{eqnarray}
\begin{eqnarray}
c^{2} &=& \frac{\left(r-\left[\frac{R_{S}}{2} + \sqrt{\left(\frac{R_{S}}{2}\right)^{2}-R_{0}^{2}e^{-2\mu r}}\right]\right)\left(
r-\left[\frac{R_{S}}{2} - \sqrt{\left(\frac{R_{S}}{2}\right)^{2}-R_{0}^{2}e^{-2\mu r}}\right]\right)}{r^{2}}c^{2}\left(\frac{dt}{d\tau}\right)^{2} \nonumber \\
&-&\frac{r^{2}}{\left(r-\left[\frac{R_{S}}{2} + \sqrt{\left(\frac{R_{S}}{2}\right)^{2}-R_{0}^{2}e^{-2\mu r}}\right]\right)\left(r
-\left[\frac{R_{S}}{2} - \sqrt{\left(\frac{R_{S}}{2}\right)^{2}-R_{0}^{2}e^{-2\mu r}}\right]\right)}\left(\frac{dr}{d\tau}\right)^{2} 
\end{eqnarray}  
Notice that the presence of charge implies that the Schwarzschild radius is not a sigularity. This implies that the linear motion 
can be realized in the Schwarzschild radius, that is close to the sigularity if the charge radius is small. The components 
$U^{2}$ and $U^{3}$ of the four 
velocity are zero.
\begin{eqnarray}
U^{0}&=&\frac{dt}{d\tau}\neq 0, \\
U^{1}&=&\frac{d r}{d\tau}\neq 0.
\end{eqnarray}
we can rewrite (\ref{89uop1}) as follows
\begin{eqnarray}
1&=&\frac{\left(r-\left[\frac{R_{S}}{2} + \sqrt{\left(\frac{R_{S}}{2}\right)^{2}-R_{0}^{2}e^{-2\mu r}}\right]\right)\left(
r-\left[\frac{R_{S}}{2} - \sqrt{\left(\frac{R_{S}}{2}\right)^{2}-R_{0}^{2}e^{-2\mu r}}\right]\right)}{r^{2}}\left(U^{0}\right)^{2}
\nonumber \\
&-& \frac{r^{2}}{\left(r-\left[\frac{R_{S}}{2} + \sqrt{\left(\frac{R_{S}}{2}\right)^{2}-R_{0}^{2}e^{-2\mu r}}\right]\right)\left(r
-\left[\frac{R_{S}}{2} - \sqrt{\left(\frac{R_{S}}{2}\right)^{2}-R_{0}^{2}e^{-2\mu r}}\right]\right)}\frac{1}{c^{2}} \left(U^{1}\right)^{2}. \label{89uop}
\end{eqnarray} 
We can then define 
\begin{eqnarray}
\cosh \lambda &=& U^{0}\sqrt{\left[\frac{\left(r-\left[\frac{R_{S}}{2} + \sqrt{\left(\frac{R_{S}}{2}\right)^{2}-R_{0}^{2}e^{-2\mu r}}\right]\right)\left(
r-\left[\frac{R_{S}}{2} - \sqrt{\left(\frac{R_{S}}{2}\right)^{2}-R_{0}^{2}e^{-2\mu r}}\right]\right)}{r^{2}}\right]} \\
\sinh \lambda &=& U^{1}\sqrt{\left[\frac{r^{2}}{\left(r-\left[\frac{R_{S}}{2} + \sqrt{\left(\frac{R_{S}}{2}\right)^{2}-R_{0}^{2}e^{-2\mu r}}\right]\right)\left(r
-\left[\frac{R_{S}}{2} - \sqrt{\left(\frac{R_{S}}{2}\right)^{2}-R_{0}^{2}e^{-2\mu r}}\right]\right)}\frac{1}{c^{2}}\right]}
\end{eqnarray}
We can then write 
\begin{eqnarray}
\tanh \lambda &=& \frac{U^{1}}{U^{0}}\frac{1}{c}\frac{r^{2}}{\left(r-\left[\frac{R_{S}}{2} + \sqrt{\left(\frac{R_{S}}{2}\right)^{2}-R_{0}^{2}e^{-2\mu r}}\right]\right)\left(r
-\left[\frac{R_{S}}{2} - \sqrt{\left(\frac{R_{S}}{2}\right)^{2}-R_{0}^{2}e^{-2\mu r}}\right]\right)}
\end{eqnarray}
On the other hand the velocity can be written
\begin{eqnarray}
v(r)= \frac{U^{1}}{U^{0}}= \frac{dr}{dt}.
\end{eqnarray}
We then have
\begin{eqnarray}
\tanh \lambda &=& \frac{v}{c}\frac{r^{2}}{\left(r-\left[\frac{R_{S}}{2} + \sqrt{\left(\frac{R_{S}}{2}\right)^{2}-R_{0}^{2}e^{-2\mu r}}\right]\right)\left(r
-\left[\frac{R_{S}}{2} - \sqrt{\left(\frac{R_{S}}{2}\right)^{2}-R_{0}^{2}e^{-2\mu r}}\right]\right)}
\end{eqnarray}
We can rewrite this expression
\begin{eqnarray}
\tanh \lambda &=& \frac{v}{c}\frac{r^{2}}{\left(r-\frac{R_{S}}{2} - \sqrt{\left(\frac{R_{S}}{2}\right)^{2}-R_{0}^{2}e^{-2\mu r}}\right)\left(r
-\frac{R_{S}}{2} + \sqrt{\left(\frac{R_{S}}{2}\right)^{2}-R_{0}^{2}e^{-2\mu r}}\right)}
\end{eqnarray}
and
\begin{eqnarray}
\tanh \lambda &=& \frac{v(r)}{c}= \frac{r^{2}}{\left(r-\frac{R_{S}}{2}\right)^{2} -\left(\frac{R_{S}}{2}\right)^{2}+ R_{0}^{2}e^{-2\mu r}}
\end{eqnarray}
As a consequence 
\begin{eqnarray}
\beta(r)= v(r)/c&=& \frac{R_{0}^{2}e^{-2\mu r}-rR_{S}}{r^{2}}\tanh \lambda(r)
\end{eqnarray}
In particular, in the case $\beta(r_{c})=1$, we have the relation, we can write
\begin{eqnarray}
\lambda (r_{c})&=& \frac{1}{2}\ln\left(\frac{R_{0}^{2}e^{-2\mu r_{c}}-r_{c}R_{S}+r^{2}_{c}}{R_{0}^{2}e^{-2\mu r_{c}}-r_{c}R_{S}-r^{2}_{c}} \right)
\end{eqnarray}
For a Schwarzschild radius, this relation reduces 
\begin{eqnarray}
\lambda (R_{S})&=& \frac{1}{2}\ln\left(\frac{R_{0}^{2}e^{-2\mu R_{S}}}{R_{0}^{2}e^{-2\mu R_{S}}-2R_{S}^{2}} \right).
\end{eqnarray}
We then have for suficiently large $r$ and for a 
\begin{eqnarray}
\lim_{r\rightarrow R>>0}\beta(R)&=& \frac{-R_{S}}{R}\tanh \lambda(R).
\end{eqnarray}
In the case of a Schwarzschild radius, we have 
\begin{eqnarray}
\beta(R_{S})&=& \left[\left(\frac{R_{0}}{R_{S}}\right)^{2}e^{-2\mu R_{S}}-1\right]\tanh \lambda(R_{S}).
\end{eqnarray}
In particular, $\beta(R_{S})=1$ will lead to a reference scaling for the Yukawa screening
\begin{eqnarray}
\mu= -\frac{1}{2R_{S}}\ln\left[\left(\frac{R_{S}}{R_{0}}\right)^{2}\left(\frac{\beta(R_{S})}{\tanh \lambda(R_{S})} + 1\right)\right].
\end{eqnarray}

\section{Tetrad fields in the proper time}

We simplify the non-coordinate basis in the following form
\begin{eqnarray}
\hat{\Theta}^{0}(r)&=& 
\sqrt{1 + \left(\frac{R_{0}}{r}\right)^{2}e^{-2\mu r} -\left(\frac{R_{S}}{r}\right)}cdt, \label{lkoi9871}\\
\hat{\Theta}^{1}(r)&=& 
\frac{1}{\sqrt{1 + \left(\frac{R_{0}}{r}\right)^{2}e^{-2\mu r} -\left(\frac{R_{S}}{r}\right)}}dr \\
\hat{\Theta}^{2}(r)&=& 
rd\theta \\
\hat{\Theta}^{3}(r,\theta)&=& 
r\sin\theta d\varphi. \label{lkoi98712}
\end{eqnarray}
In the proper time, the non-coordinate basis are reduced to
\begin{eqnarray}
\hat{\Theta}^{0}(\tau,r)&=&  
\sqrt{1 + \left(\frac{R_{0}}{r}\right)^{2}e^{-2\mu r} -\left(\frac{R_{S}}{r}\right)}cU^{0}d\tau, \\
\hat{\Theta}^{1}(\tau,r)&=& 
\frac{1}{\sqrt{1 + \left(\frac{R_{0}}{r}\right)^{2}e^{-2\mu r} -\left(\frac{R_{S}}{r}\right)}} U^{1}d\tau
\end{eqnarray}
We can then write
\begin{eqnarray}
\frac{\hat{\Theta}^{1}(\tau,r)}{\hat{\Theta}^{0}(\tau,r)}&=& 
\frac{\beta(r)}{1 + \left(\frac{R_{0}}{r}\right)^{2}e^{-2\mu r} -\left(\frac{R_{S}}{r}\right)}  d\tau
\end{eqnarray}
We can also write
\begin{eqnarray}
c^{2}d\tau^{2}= \hat{\Theta}^{0}(\tau,r)\hat{\Theta}^{0}(\tau,r) - \hat{\Theta}^{1}(\tau,r)\hat{\Theta}^{1}(\tau,r)
\end{eqnarray}
We then have
\begin{eqnarray}
\frac{c^{2}d\tau^{2}}{\hat{\Theta}^{0}(\tau,r)\hat{\Theta}^{0}(\tau,r)}= 
1 - \frac{\hat{\Theta}^{1}(\tau,r)\hat{\Theta}^{1}(\tau,r)}{\hat{\Theta}^{0}(\tau,r)\hat{\Theta}^{0}(\tau,r)}
\end{eqnarray}
and then
\begin{eqnarray}
(U^{0}(r))^{-1}= 1 + \left(\frac{R_{0}}{r}\right)^{2}e^{-2\mu r} -\left(\frac{R_{S}}{r}\right) 
- \frac{\beta(r)^{2}}{\left[1 + \left(\frac{R_{0}}{r}\right)^{2}e^{-2\mu r} -\left(\frac{R_{S}}{r}\right)\right]}.
\end{eqnarray}
We then
\begin{eqnarray}
(U^{0}(R_{S}))^{-1}e^{-2\mu R_{S}}= \left(\frac{R_{0}}{R_{S}}\right)^{2}e^{-4\mu R_{S}}
- \frac{\beta(R_{S})^{2}}{\left(\frac{R_{0}}{R_{S}}\right)^{2}}.
\end{eqnarray}
We can then also write the screening scale in the following form
\begin{eqnarray}
\mu_{\pm}&=& -\frac{1}{2R_{S}}\ln\left[\frac{(U^{0}(R_{S}))^{-1} \pm \sqrt{(U^{0}(R_{S}))^{-2} + 4\beta(R_{S})^{2}}}{2\left(\frac{R_{0}}{R_{S}}\right)^{2}}\right].
\end{eqnarray}
In this case, the screening behaviour is due to $\mu_{+}$ while $\mu_{-}$ is a complex solution.  

\section{Conclusions}

The Reissner-Nordstrom spacetime is an important extension of the Schwarzschild spacetime, where it is considered an aditional 
contribution of charge. A Yukawa-type electric field implies that the charge is screened at a large distance and the electric field has a 
finite range. As a consequence, the Yukawa-Reissner-Nordstr\"om spacetime can be used for studying the singularity transition from a Reissner-Nordstr\"om 
spacetime to a Schwarzschild spacetime, with reduction of the double singularites at large scale. 

We considered the tetrad fields associated to Yukawa-Reissner-Nordstr\"om spacetime and gave estimatives of the Yukawa screening parameter in the Schwarzschild 
radius in the cases of linear motion in the proper time under constant angular velocities 
and considering the tetrad fields in the proper time.

Implications in the cosmic censorship hypothesis and the behaviour of naked sigularities can be studied in the
 scenario of Yukawa-Reissner-Nordstr\"om spacetime. This also can lead to another route in the case of extremal black holes, as the Yukawa screening 
 of charge lead to a 
 correction in a extremal black hole relation in \cite{wilczek}, implying a screening in the in the Mass-Charge relation 
 \begin{eqnarray}
 M^{2}(r)=\frac{Q_{0}^{2}e^{-2\mu r}}{1+a^{2}}.
 \end{eqnarray}
In this case, the singularity can be viewed as a dynamical entity as a result of transition from a Reissner-Nordstr\"om to a Schwarzschild singularity 
in the presence of a Yukawa screening.

\section{Acknowledgements}

The author thanks the support by FAPEMA (Brazil)- APCINTER-00273/14, Chamada Interna Enxoval – UFMA PPPG N03/2014 (Brazil), and institutionalized projects UFMA-Res.No1342-CONSEPE Art1-III-1150/2015-33, 
UFMA-Res.No1342-CONSEPE Art1-IV-1151/2015-88.

\end{widetext}

\end{document}